\def\cm2{cm$^{-2}$}

\def\c2{C~{\sc ii}}
\def\c4{C~{\sc iv}}
\def\fe2{Fe~{\sc ii}}
\def\fe3{Fe~{\sc iii}}
\def\mg1{Mg~{\sc i}}
\def\mg2{Mg~{\sc ii}}
\def\si2{Si~{\sc ii}}
\def\si4{Si~{\sc iv}}
\def\al2{Al~{\sc ii}}
\def\al3{Al~{\sc iii}}
\def\o1{O~{\sc i}}
\def\n1{N~{\sc i}}
\def\h1{H~{\sc i}}

\def\approxlt{\mathrel{\spose{\lower 3pt\hbox{$\sim$}}
        \raise 2.0pt\hbox{$<$}}}
\def\approxgt{\mathrel{\spose{\lower 3pt\hbox{$\sim$}}
        \raise 2.0pt\hbox{$>$}}}

\documentclass[article]{gwp80}
\usepackage{graphicx}
\usepackage{amssymb}
\tabletypesize{\normalsize}  

\def\plotone#1{\centering \leavevmode
\includegraphics[width=.95\columnwidth]{#1}}

\def\plotone#1{\centering \leavevmode
\includegraphics[width=.95\columnwidth]{#1}}

\shortauthors{Breger}
\shorttitle{Rotational modulation in RRs/$\delta$ Scuti star pulsation}

\begin{document}
\large    
\pagenumbering{arabic}
\setcounter{page}{101}

\title{Rotational modulation in RRs/$\delta$ Scuti star\\pulsation}

%
%
\author{{\noindent Michel Breger{$^{\rm 1,2}$}}\\
\\
{\it (1) Institut f\"ur Astronomie der Universit\"at Wien, T\"urkenschanzstr. 17, Wien, Austria\\
(2) Department of Astronomy, University of Texas, Austin, TX 78712, USA} 
}

%
%
\email{(1) breger@astro.as.utexas.edu}


\begin{abstract}
The 76 frequencies determined from $Kepler$ measurements of the A-type star KIC 9700322 can be understood in a rather simple way as the radial fundamental and first overtone radial modes, a nonradial $\ell=2$ quintuplet, stellar rotation and combinations of all of these. Two low-frequency peaks are found: f=0.160 c$\,$d$^{-1}$ and a 2f peak. We show from a variety of arguments that the low-frequency peak, f, (rather than the 2f peak) is the rotational frequency.
The corresponding prediction of slow rotation is confirmed by a spectrum from which $v \sin i=19 \pm 1$\,km\,s$^{-1}$ is obtained.

During rotation, the brightness changes by 0.5 ppt (parts-per-thousand, peak-to-peak). The rotation also modulates the amplitudes of the two radial modes, leading to a small (but definite) variation in amplitude of 0.5$\%$ of their average values. Both the rotational brightness variations and the pulsation amplitudes form a double wave during rotation. The variations are in phase. The maximum pulsation amplitude occurs at the rotational phase with minimum brightness. Consequently, the 'normal' star KIC 9700322 behaves such as an extremely weak chemically peculiar star. Since such variations are also seen in some other A stars studied by $Kepler$, normal stars with a symmetric surface may not exist.

The significance of this star is that even a simple, slowly rotating, radial pulsator already shows rotational brightness variations and modulation of the pulsation amplitude.

 \end{abstract}

\section{Introduction}

In 1966 George W. Preston suggested a dissertation topic on the essentially unknown group of the
so-called RRs stars to a new Ph.D. student (the author of this paper) and guided the
student through the new field of millimag-accuracy photometry at Lick Observatory. The RRs stars later became known
as Delta Scuti stars, in particular the high-amplitude Delta Scuti stars (HADS). These resemble
RR Lyrae stars in many ways, such as dominant radial pulsation.
The millimag accuracy has now given way to an accuracy measured
in parts-per-million (ppm) obtained with space satellites.

This paper is based on measurements of the star KIC 9700322 (Breger et al. 2011)
obtained with the $Kepler$ satellite (Koch 2010). In the present paper we briefly
examine the simplicity of the pulsational properties of this star (despite the 76+ detected
frequencies) and highlight the effects of rotation in this normal star.

A spectrum obtained with the McDonald Observatory 9.2 m HET telescope reveals
that the star is one of the coolest
$\delta$\,Scuti variables with a temperature of  T$_{\rm eff}$\,=\,6700~$\pm$~100 K and
$\log g$\,=\,3.7~$\pm$~0.1. The abundances are found to be solar. Furthermore,
the star is found to be sharp-lined with a measured value $v \sin i=19 \pm 1$\,km\,s$^{-1}$ .

\begin{figure*}
\centering
\plotone{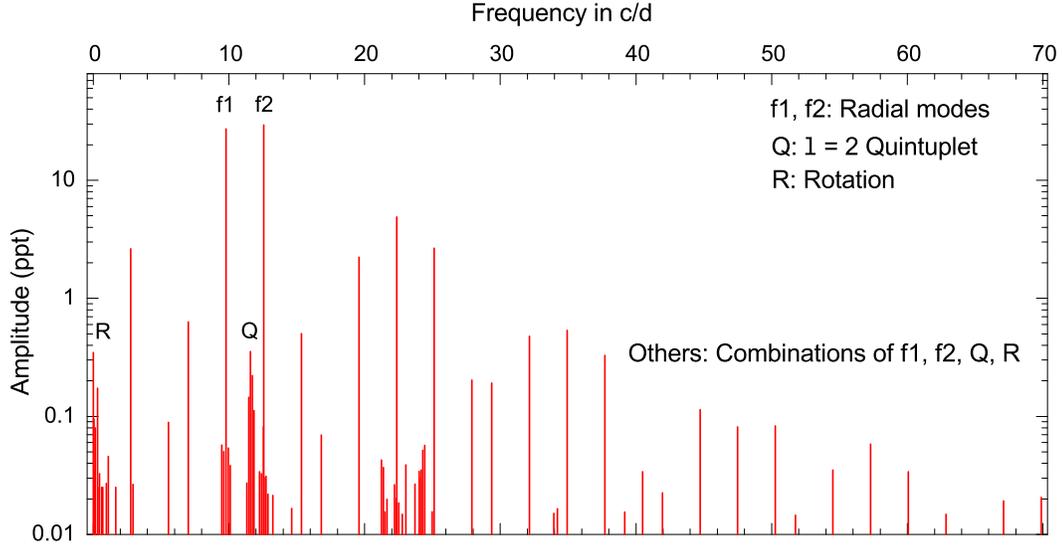}
\vskip0pt
\caption{The detected amplitude spectrum of KIC 9700322. The amplitudes shown cover a range of more than a factor of 1000. The
frequency patterns are quite simple and can be described by four physical features (see text).}
\label{rotn}
\end{figure*}

\section{Pulsational properties of KIC 9700322}

The short-cadence Q3 {\it Kepler} data of KIC\,9700322, covering 30.2 d, were analyzed with the statistical package {\tt 
PERIOD04} (Lenz \& Breger 2005). This package carries out multifrequency analyses 
with Fourier as well as least-squares algorithms and does not rely on the 
assumption of white noise. After prewhitening 76 statistically significant frequencies, 
the remaining noise levels in the Fourier diagrams ranged from 0.007\,ppt at frequencies
below 10 c$\,$d$^{-1}$ to 0.0036\,ppt in the 40 - 200 c$\,$d$^{-1}$ range.

An unexpected result is the absence of hundreds of frequency peaks with amplitudes in excess of the low limit of 0.015 ppt.
This behavior is also shown by several other $\delta$ Scuti stars observed by the $Kepler$ satellite. Even more surprising
is the relative simplicity of the frequency spectrum, since the majority of peaks are combination frequencies.
The relative dominance of the combination peaks is also a feature found in other pulsators studied by $Kepler$.

The amplitude spectrum of KIC 9700322 is shown Fig. 1. The spectrum is dominated by four features:

(i) The two dominant modes (f$_1$, f$_2$) can be identified with the radial fundamental and first overtone modes, which
is typical for the sharp-lined HADS (High-Amplitude Delta Scuti) group.
\footnote {A popular definition of the HADS (RRs) group also includes a peak-to-peak amplitude
limit of 0.3 mag, which is not reached by KIC 9700322. However, a clean separation of the $\delta$ Scuti variables into completely separate
nonradial low-amplitude and radial high-amplitude pulsator groups is too simple. A diagram of rotational velocity vs. amplitude indicates
a transition between the two groups (see Breger 2000). The presently available data indicate that a
necessary condition for dominant radial pulsation in $\delta$ Scuti stars is slow rotation.}

(ii) A nonradial $\ell$=2 quintuplet with an average frequency separation of 0.134\,\nolinebreak c$\,$d$^{-1}$ (named Q in the figure).

(iii) Two low-frequency peaks near 0.160 and 0.320 c$\,$d$^{-1}$ (named R in the figure). We shall demonstrate below that
they are connected to the rotational frequency.

(iv) A large number of combination frequencies of the frequencies mentioned under (i) to (iii). This includes combinations of separate frequencies,
other nonlinearities caused by Fourier series of a nonsinuoidal light curve and the amplitude modulations during rotation. \footnote {If one examines the
frequency regularities in KIC 9700322 (and many other $\delta$ Scuti stars studied with the $Kepler$ satellite), the dominant
frequency patterns are caused by combination modes, $nf_i \pm mf_j$, where $m$ and $n$ are integers. One of the most common
(and dominant to the eye searching for patterns) is the basic difference in two frequencies involved in the combinations, viz. $f_i-f_j$.
For KIC 9700322 this is 2.7763\,cd$^{-1}$. This value is the dominant frequency separation in Fig. 1. Also, the combination-mode peak at 2.7763\,cd$^{-1}$
should not be confused with a gravity mode.}

\begin{figure*}
\centering
\plotone{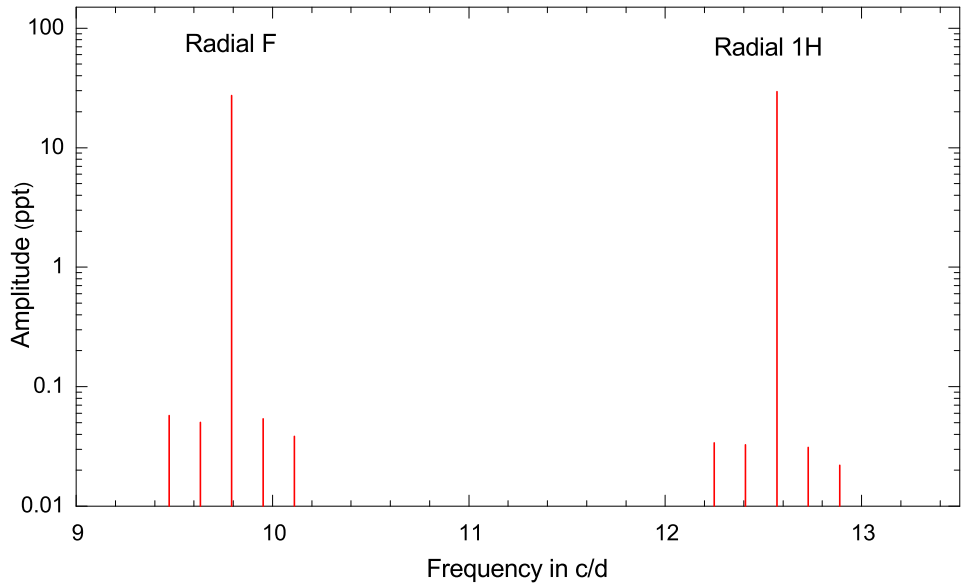}
\vskip0pt
\caption{Sidelobes of the two radial modes in the Fourier domain.
The absolutely equidistant sidelobes are separated by exactly the rotational frequency.
The behavior can be explained by rotational modulation of the amplitudes.}
\end{figure*}

\begin{figure*}
\centering
\plotone{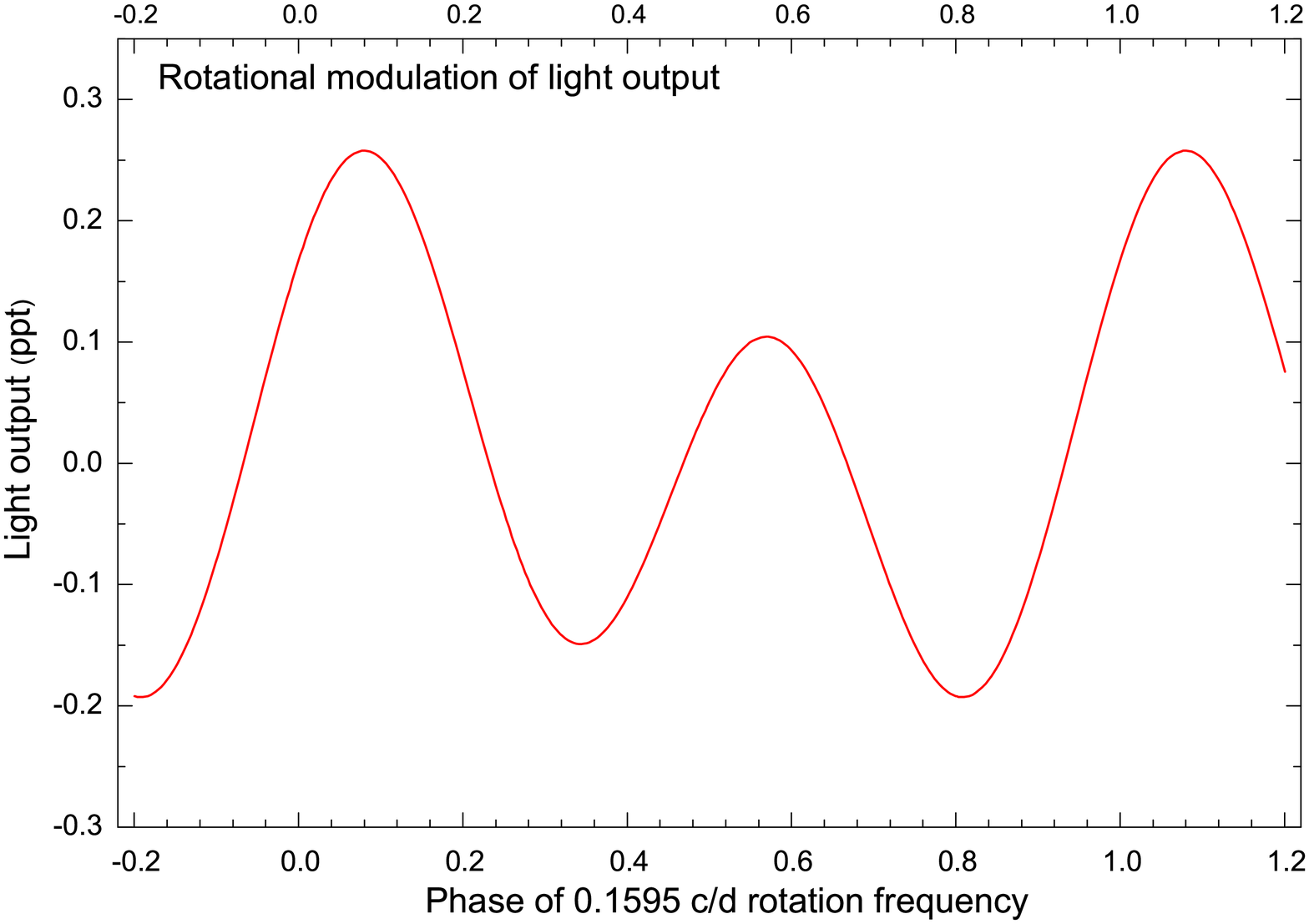}
\vskip0pt
\caption{Variation of the stellar brightness with rotation in the Q3 $\it{Kepler}$ data.
To obtain this curve, the pulsation modes with their detected combinations and modulations were
prewhitened (see Breger et al. 2011). The double-wave rotation curve is stable over the 30.2 days. Zero phase
was chosen to be same as in Fig. 4.}
\end{figure*}

What are the arguments in favor of identifying the 0.160 c$\,$d$^{-1}$ peak with the rotation frequency?

(i) The fitting of the observed line profiles to model atmospheres yields T$_{\rm eff}$\,=\,6700~$\pm$~100 K and
$\log g$\,=\,3.7~$\pm$~0.1. The 0.160 c$\,$d$^{-1}$ peak predicts a rotational velocity near 23 \,km\,s$^{-1}$, in agreement with
the measured value of $v \sin i$=19 $\pm 1$\,km\,s$^{-1}$.

(ii) The average spacing of the $\ell$ = 2 quintuplet is 0.134 c$\,$d$^{-1}$. 
After applying the appropriate Ledoux Constant for the rotational splitting computed from pulsation models for the $\ell$ = 2 modes,
we predict a rotation frequency of 0.160 c$\,$d$^{-1}$, which is exactly as observed.

(iii) Fig. 2 shows the Fourier spectrum near the frequencies of the radial modes. In both cases sidelobes separated by exactly 0.160 c$\,$d$^{-1}$ appear.
An interpretation in terms of $\ell$ = 2 quintuplets (and misidentification of the radial nature of the dominant peaks) is ruled out by
the equidistant spacing of the peaks and the fact that these patterns with this exact spacing are repeated at many different frequencies.
The features can be explained by amplitude modulation. The measured separation of 0.1597 $\pm$ 0.0002 c$\,$d$^{-1}$ agrees with the values obtained above.

\section{Rotational modulation of the light output}

Let us examine the rotational modulation of the light output in more detail. In order to derive the average rotation light curve, we have determined the
best 76-frequency fit for the 30.2d of data and prewhitened the 72 frequencies, omitting the rotation frequency and its three multiples.
This leaves the rotation peaks as the dominant source of variability in the data. (We note that the 2f
multiple is close to two spurious periods near 3 d caused by instrumental problems of $Kepler$ data: consequently, the amplitude of the
2f term may be uncertain.) The resulting average light curve is shown in Fig. 3.
We see that during rotation the light output changes with a small peak-to-peak amplitude of 0.5\%. The light
curve forms a double wave. The variations are reminiscent of those of chemically peculiar and magnetic stars. However, the available data for
KIC 9700322 suggest that it is a normal, rather than a chemically peculiar star.

\section{Rotational modulation of the pulsation amplitudes}

\begin{figure*}
\centering
\plotone{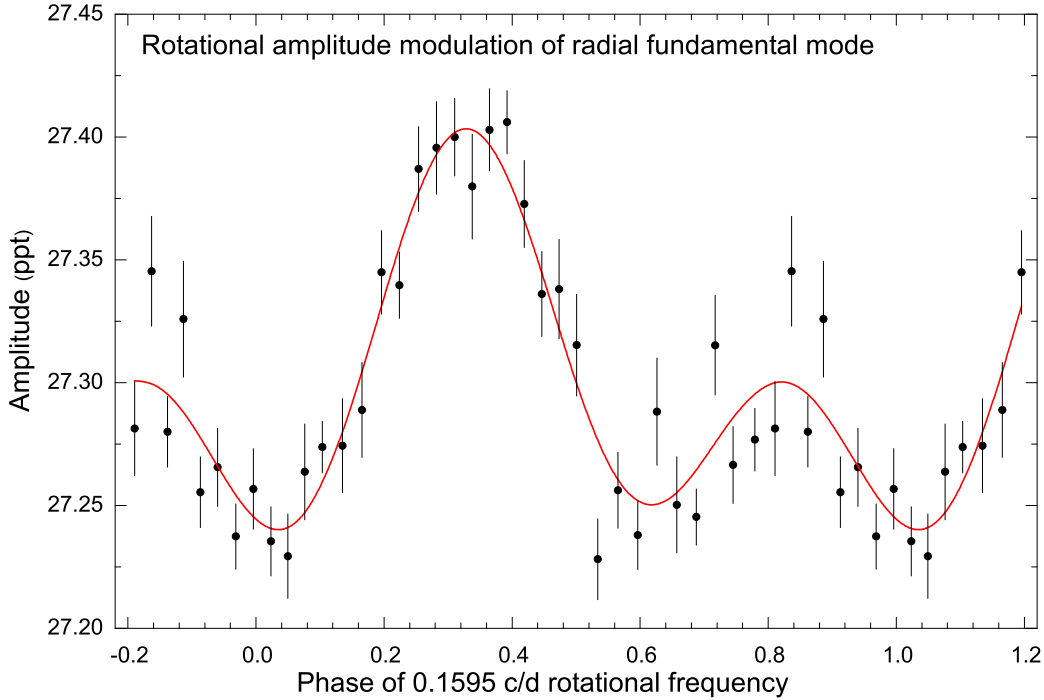}
\vskip0pt
\caption{Variation of the amplitude of the radial fundamental mode with rotation, derived from the Q2 and Q3 data. Each point shown is an average of 10 amplitudes deduced from cycles of similar rotation phases.
Note the small, but systematic change with rotation and the double wave. The amplitudes of the radial first overtone mode show similar changes during rotation.}
\label{rotnf1}
\end{figure*}

The pulsation amplitudes of both radial modes change during rotation, as already hinted at by the presence of equidistant sidelobes in the Fourier domain (Fig. 2).
In order to examine these changes, we have determined the best 76-frequency fit and prewhitened all frequencies except for the chosen radial mode with its sidelobes.
We then computed the amplitude for each individual pulsation cycle and phased it with the period of rotation. The large scatter in the resulting amplitude curve as a function
of rotational phase was reduced by forming groups of 10 amplitude values at similar rotation phases. This also allowed us to compute the uncertainty of each average value.
The results are shown for the radial fundamental mode in Fig. 4. The curve for the radial first overtone mode looks similar.

The fit of the double wave is excellent and the scatter of the individual points is in agreement
with the values expected from the error bars. The amplitudes of the radial modes exhibit a peak-to-peak variation of 0.15 ppt during rotation and are only  0.5\% of the amplitude values.
Such small changes would not have been possible to determine with standard earth-based photometry.

We conclude that both the light output and the amplitudes of the radial modes change systematically during rotation.
In fact, these two variations are not independent of each other. A comparison of Figs. 3 and 4, which have the same phase zero-point, indicates that
the two variations are in phase: the maximum pulsation amplitude occurs at the rotational phase with minimum light output, and vice versa.

\section{Comparison of this normal star with a magnetic chemically peculiar star}

In the previous sections we have seen that the brightness of KIC 9700322 and the amplitudes of pulsation vary with rotation. This behavior is typical for chemically
peculiar stars. The size of these effects in KIC 9700322 is very small (but significant).

Recently, the roAp star, $\alpha$ Cir, was studied in great detail by the $WIRE$ satellite (Bruntt et al. 2009). In Table 1 we compare the
properties of this chemically peculiar star with those of KIC 9700322. This table illustrates that we can exclude the hypothesis that KIC 9700322 is a roAp star because of the pulsation periods and abundances.

We note that many other A stars studied with $Kepler$ also show the luminosity variation with rotation. This shows up as two low-frequency peaks (f, 2f) in the Fourier diagram.
In this paper we have shown that in KIC 9700322 it is the f (rather than the 2f) peak which corresponds to the rotation frequency. These peaks indicate that the surfaces of these stars are not
homogeneous and that (large) spots may exist. Could it be that $Kepler$ has revealed that there exists a continuous range between 'normal' and peculiar stars?

\begin{flushleft}
\begin{deluxetable*}{lll}
\tabletypesize{\normalsize}
\tablecaption{Comparison of $\alpha$ Cir with KIC 9700322}
\tablewidth{0pt}                                                       
\tablehead{  & \multicolumn{1}{l}{$\alpha$ Cir}  &  \multicolumn{1}{l}{KIC 9700322} }      
\startdata

Type	 & roAp & Normal $\delta$ Scuti (HADS)\\
T$_{\rm eff}$ & 7400 K &	6700 K \\
Rotation period	 &			4.5 d		&			6.3 d\\
Rotational modulation &	4 ppt		&			0.5 ppt\\
(peak-to-peak) &					(double wave)	&		(double wave)\\
Main pulsation periods &		211 cd$^{-1}$	&		9 to 13 cd$^{-1}$\\
Pulsation amplitude	&		1.5 ppt ($\ell$ = 1)	&	27 + 29 ppt (F + 1H radial modes)\\
& 				&							0.03 to 0.4 ppt ($\ell$ = 2 quintuplet)\\
Magnetic field & 239 $\pm$ 27 Gauss	& unknown\\		
Abundances	&			Co, Y, Nd and Eu + 1 dex & normal\\
\enddata
\end{deluxetable*}
\end{flushleft}	

\acknowledgements
This investigation has been supported by the Austrian Fonds zur F\"{o}rderung der wissenschaftlichen Forschung through project P 21830-N16.

\end{document}